\documentclass[journal=jpcbfk,manuscript=article]{achemso}

\usepackage{amsmath}
\usepackage{amssymb}
\usepackage{units}
\usepackage{color}
\usepackage{graphicx}
\usepackage{dcolumn}
\usepackage{bm}
\usepackage{fixmath}
\usepackage{float}
\usepackage{upgreek}

\def\lsop#1{{\boldsymbol{\mathcal{#1}}}}
\def\lsopNonCal#1{{\boldsymbol{#1}}}

\title{Strong Exciton-Vibrational Coupling in Molecular Assemblies. Dynamics using the Polaron Transformation in HEOM Space}

\author{Joachim Seibt}
\affiliation{%
Institute of Physics, University of Rostock, Albert-Einstein-Str. 23-24, 18059 Rostock, Germany
}%
\altaffiliation{%
Institute for Theoretical Physics, Johannes Kepler University Linz, Altenberger Str. 69, 4040 Linz, Austria
}%

\email{joachim.seibt@jku.at}
\author{Oliver K\"{u}hn}%
\affiliation{%
Institute of Physics, University of Rostock, Albert-Einstein-Str. 23-24, 18059 Rostock, Germany
}%

\begin{document}
%

\begin{abstract}
In the context of Frenkel exciton dynamics in aggregated molecules the polaron transformation technique facilitates a treatment where diagonal elements attributed to electronic excited-state populations are decoupled from fluctuations associated with vibrational degrees-of-freedom. In this article we describe for the first time how the polaron transformation can be applied in the context of the ``Hierarchical Equations of Motion'' (HEOM) technique for treatment of open quantum systems with all vibrational components attributed to an environment. By using a generating function approach to introduce a shift in the excited state potential energy surface, we derive hierarchical equations for polaron transformation in analogy to those for time propagation. We demonstrate the applicability of the developed approach by calculating the dynamics of underdamped and overdamped oscillators coupled to electronic excitation of a monomer without and with previous polaron transformation and study the dynamics of the expectation value of the respective vibrational coordinates. Furthermore, we investigate the dynamics of a dimer with a barrier comparable to the thermal energy between the minima of the lower excitonic potential energy surface. It turns out that the assumption of localization at the monomer unit with energetically higher potential minimum, introduced via polaron transformation, has a substantial influence on the transfer dynamics. Here, it makes a clear difference whether the polaron transformation is performed in the local or exciton basis. This reflects the fact that the polaron transformation only accounts for equilibration of the vibrational, but not of the excitonic dynamics. We sketch an approach to compensate this shortcoming in view of obtaining an initial state for the calculation of emission spectra of molecular aggregates.
\end{abstract} 

\maketitle

\newpage
\section{Introduction}
In molecular aggregates, photosynthetic pigment-protein complexes and organic materials electronic exitations are coupled to  vibrational degrees of freedom (DOF), such that their dynamics cannot be disentangled \cite{May11,valkunas13,renger01_137,BaBuVa14_JCP_95,schroter15_1,kuhn18_259,gosh20,meggiolaro20}. In analogy to a transformation to the exciton basis, where diagonalization of the purely electronic part of the Hamiltonian turns the off-diagonal Coulomb coupling into contributions to the exciton eigenenergies, it is also possible to apply a transformation to the vibrational parts of the Hamiltonian, such that the system-bath coupling -- in the general sense of any couplings of vibrational modes to electronic excitations -- vanishes. Such transformation, which is obtained by applying a shift operator \cite{May11,BaBuVa14_JCP_95} to compensate the displacement of the equilibrium position in excited-state potentials of vibrational modes compared to the ground-state potential, is known as the polaron transformation (PT)  \cite{SiHa84_JCP_2615}. Apart from applications where a quantum mechanical treatment with involvement of vibrational eigenstates was chosen \cite{BlSiSt16_CP_250}, it has been extensively applied in the context of open quantum systems in the framework of spin-boson models \cite{ChPrHu11_PRL_160601} with applications ranging from quantum dots \cite{ChStGo20_PRB_035306},  molecular donor-acceptor complexes \cite{YaDeJa12_JCP_024101}, interacting excitonic dimer units \cite{schroter15_536,MaMoCa15_JCP_094197} and light-harvesting complexes \cite{KoNaOl11_JCP_154112,PoMcLo13_NJP_075018} up to bulk materials, such as organic molecular crystals \cite{ChYeZh11_JPCB_5312,ZhBrLi94_JCP_2335,MuSi85_JCP_1843,WaZh19_JCompChem_1097}.
In particular, second-order perturbative description of transfer processes in molecular aggregates with separation of reference and interaction Hamiltonian via PT is a widely used application. It has mostly been formulated in the localized basis with basis states corresponding to electronic excitation of a single monomer unit \cite{SiHa84_JCP_2615,LeMoCa12_JCP_204120,SuFuIs16_JCP_204106,YaDeJa12_JCP_024101,ChZhCh13_JCP_224112}, but also in the exciton basis \cite{XuCa16_FrontPhys_110308,XuWaYaCa16_NJP_2016,Ki16_JCP_123,Ki16_JCP_70,HsLiDu19_JCP_17196}. Such second-order perturbative treatment goes beyond the assumptions entering in F\"{o}rster- or Redfield type approaches that either the excitonic coupling or the system-bath coupling is sufficiently small to be treated perturbatively \cite{YaFl02_CP_355,SeMa17_JCP_174109,SeKu20_JCP}. It is therefore also applicable in cases where neither of them is appropriate, for example for the treatment of model systems with off-diagonal contributions to the system-bath coupling \cite{SuFuIs16_JCP_204106}.
While in a quantum mechanical treatment with vibrational eigenstates or in an open quantum system description with formulation of a quantum master equation (QME) for the reduced density matrix the coupling of vibrational degrees of freedom to electronic excitations enters in terms of off-diagonal elements or via the correlation function, respectively, in the framework of ``Hierarchical Equations of Motion'' (HEOM) it seems not to be obvious at first glance how to identify the system-bath coupling and how to apply a shift operator for polaron transformation. 

The HEOM approach was pioneered by Y. Tanimura and co-workers~\cite{tanimura89_101,Ishizaki05JPCJ,Tanimura06JPCJ,Tanimura20_JCP_020901} and has been developed into a standard for nonperturbative and non-Markovian calculations \cite{Tanimura20_JCP_020901,ShChNa09_JCP_164518,ShChNa09_JCP_084105,Chen09JCP,Chen10JCP,KrKr12_JPCL_2828,HeKrKr12_NJP_023018,StSc12_JCTC_2808,LiZhBa14_JCP_134106,OlKr14_JCP_164109,WiDa15_JCTC_3411,schroter15_536,XuSo17_JCP_064102,DiPr17_JCP_064102}. In the HEOM method all nuclear DOF are attributed to the environment \cite{Ta12_JCP_22A550}, and all orders of perturbation theory with respect to their interaction with the system built from purely electronic basis states are taken into account, at least in principle. However, in view of analogies between HEOM and description in a vibronic basis, which rely on the possibility to interpret the Auxiliary Density Operators (ADOs) of the Kubo-Tanimura hierarchy \cite{Tanimura06JPCJ} as a representation of a (stochastic) vibrational coordinate \cite{ShChNa09_JCP_164518,LiZhBa14_JCP_134106}, the applicability of the concept of polaron transformation in the context of HEOM becomes comprehensible. Recently, an approach has been proposed for taking dependencies of excitonic couplings or transition dipole moments on vibrational modes into account at the level of the hierarchical equations of motion \cite{SeMa18_CP_129}. Here, we are aiming at an analogous approach for the polaron transformation by expressing the shift operator in a differential form to obtain hierarchical equations for a polaron transformation in HEOM space (i.e.\ in the space where the ADOs are identified with vector components), which we can then integrate up to a given displacement or to an arbitrary position resulting from variational approaches \cite{SuFuIs16_JCP_204106,YaDeJa12_JCP_024101} in the context of second-order rate theories. 

The PT has not been discussed in the context of HEOM so far, apart from a reference  related to the calculation of second-order transfer rates \cite{SeKu20_JCP}. There, the applicability of PT to account for thermal equilibration in an excited initial state of the transfer process has been demonstrated and the application of PT to separate reference and interaction Hamiltonian in the framework of a variational QME, where the shift resulting from PT does not necessarily correspond to the excited state displacement, has been sketched. Moreover, according to Ref. \citenum{XuCa16_FrontPhys_110308} a steady state can be determined from a formulation of the latter approach in the exciton basis. As such a steady state is thermally equilibrated with respect to both vibrational and exciton dynamics, its determination in the context of HEOM would be useful to obtain an initial state for the calculation of emission spectra of molecular aggregates with HEOM, as a complement to the approach proposed in Ref. \citenum{JiChBa13_JCP_045101}.

This article is organized as follows: In the next section we describe the theoretical background. Starting from a specification of the exciton-vibrational Hamiltonian and a general formulation of HEOM, we derive analogous hierarchical equations for the PT. Furthermore, we describe how expectation values of vibrational coordinates are calculated from the ADOs and how absorption and emission spectra are obtained from HEOM calculations. Afterwards, we first demonstrate the PT in HEOM space by comparing the dynamics without and with polaron transformation for an underdamped (Brownian) oscillator and for an overdamped bath characterized by a Debye-Drude spectral density. To provide further evidence for the correctness of the PT, we asses whether the expected mirror symmetry of monomer absorption and emission spectra is obtained when the PT enters in the calculation of the latter to account for the assumption of initial equilibration in the excited state. The main part of the discussion of our results is related to the excited state dynamics of a dimer with parameters chosen in such way that along the antisymmetic linear combination of the monomer vibrational coordinates a double minimum structure appears in the energetically lower excitonic potential. The barrier to overcome for getting from the energetically higher to the energetically lower local minimum is adjusted to be of the order of the thermal energy. We study the cases without and with PT before the propagation and explain the differences in the population dynamics and in the time evolution of the expectation values of the vibrational coordinates. We further discuss the effect of performing the PT in the local and exciton basis and relate it to the only partial (vibrational) equilibration achieved by this transformation.
As an outlook we sketch how a thermally equilibrated state with respect to both vibrational and excitonic dynamics can be obtained by in the context of HEOM by assuming a steady state for the second-order rate equation with involvement of a PT.
Finally, we summarize our results, draw conclusions and point to open questions. 
%
\section{Theoretical Background} \label{sec:theoretical_background}
\subsection{Exciton-Vibrational Hamiltonian}
%
The exciton-vibrational Hamiltonian to be used in the following is rather standard and of the general system-bath form
$\hat{H}=\hat{H}_{\rm S}+\hat{H}_{\rm B}+\hat{H}_{\rm SB}$~\cite{May11,schroter15_1}. 
The (excitonic) system part is given by 
\begin{equation} \label{eq:Hamiltonian_general_system}
\hat{H}_{\rm S}=\sum_{lm} ( \delta_{lm} (\epsilon_l + \lambda_l) + J_{lm} )\hat{B}^{\dagger}_l \hat{B}_m \,. 
\end{equation}
Here, $\epsilon_l$ and $J_{lm}$ are site energies and Coulomb coupling, respectively, and we further introduced the exciton creation, $\hat{B}^{\dagger}_l$, and annihilation, $ \hat{B}_l$, operators. The state $l$ implies that the $l$-th monomer is in the electronically excited state $e_l$, whereas all other monomers are in the ground state $g_{m\ne l}$.  

Two different models concerning the system-bath, i.e. exciton-vibrational, coupling will be considered (for a general classification, see Ref. \citenum{schroter15_1}).
In the first model, the coupling is to a thermal bath described by (phonon) coordinates $\{q_i\}$ and frequencies  $\{\omega_i\}$  ($\hbar=1$)
\begin{equation}
  \label{eq:Hamiltonian_general_bath}
\hat{H}_{\mathrm B} = \frac{1}{2}\sum_i \Big( \hat{p}_i^2 + \omega_{i}^2 \hat q_{i}^2 \Big)  = \sum_i\frac{\omega_{i}^2}{2} \Big(-\frac{\partial^2}{\partial q_{i}^2} + \hat q_{i}^2 \Big    ) \, .
\end{equation}
Note that, different from Ref. \citenum{SeKu20_JCP}, the coordinates introduced for the definition of the contributions to the Hamiltonian are not rescaled to become dimensionless.
Momentum and position operator can be expressed in terms of bosonic creation and annihilation operators as $\hat{p}_i=i \sqrt{{\omega_i}/{2}}(\hat{b}_i^{\dagger}-\hat{b}_i)$ and $\hat{q}_i=\sqrt{{1}/{2 \omega_i}}(\hat{b}_i^{\dagger}+\hat{b}_i)$.
The coupling of phonon modes to electronic transitions is commonly described by the displaced oscillator model with the displacement $d_{l,i}$ connected to the Huang-Rhys factor $S_{l,i}$ and the coupling $g_{l,i}=\sqrt{S_{l,i}}$ in terms of $d_{l,i}=\sqrt{{2 S_{l,i}}/{\omega_{i}}}=g_{l,i} \sqrt{2/{\omega_{i}}}$, leading to the system-bath coupling Hamiltonian
%
%
\begin{equation}
\label{eq:HSB1}
\begin{split}
 \hat{H}_{\rm SB}=\sum_l \hat{H}_{{\rm SB},l} &= -\sum_{l,i}  d_{l,i} \omega_{i}^2 \hat q_i \hat{B}^{\dagger}_l \hat{B}_l \\
 &= -\sum_{l,i}  \sqrt{S_{l,i}} \omega_{i} (\hat{b}_i^{\dagger}+\hat{b}_i) \hat{B}^{\dagger}_l \hat{B}_l \\
 &= -\sum_{l,i}  g_{l,i} \omega_{i} (\hat{b}_i^{\dagger}+\hat{b}_i) \hat{B}^{\dagger}_l \hat{B}_l.
 \end{split}
\end{equation}
For this model the total reorganization energy at monomer $l$ entering Eq. \eqref{eq:Hamiltonian_general_system} is given by $\lambda_l=\sum_i \omega_i S_{l,i}$.

 The actual distribution of couplings $\{g_{l,i}\}$ will be described by a Debye-Drude (DD) spectral density, which will be taken to be equal for all sites (skipping the site index) 
\begin{equation} \label{eq:Debye-Drude_spectral_density}
J_{\rm DD}(\omega)= 2 \lambda_{\rm DD} \frac{\omega_{\rm c} \omega}{\omega^2+\omega_{\rm c}^2}
\end{equation}
Here, $1/\omega_{\rm c}$ and  $\lambda_{\rm DD}$  are bath correlation time and  bath reorganization energy upon electronic excitation, respectively~\cite{Mukamel95}.

In the second system-bath model, we will describe the case of coupling to a damped high-frequency (intramolecular) mode. This is commonly done using the multimode Brownian oscillator (BO) model, which assumes a bilinear coupling to both electronic excitation and an additional thermal bath. In the Ohmic dissipation limit the influence of the thermal bath can be treated in terms of a phenomenological damping constant.
Then the bath Hamiltonian and the system-bath coupling Hamiltonian of the BO can be formulated in analogy to Eqs.~(\ref{eq:Hamiltonian_general_bath}) and (\ref{eq:HSB1}).
The spectral density for the BO model in case of a single mode with frequency $\omega_{\rm BO}$, Huang-Rhys factor $S_{\rm BO}$, reorganization energy $\lambda_{\rm BO}=S_{\rm BO} \omega_{\rm BO}$, and damping $\gamma_{\rm BO}$ is given by~\cite{Mukamel95}
\begin{equation} \label{eq:Brownian_oscillator_spectral_density}
J_{\rm BO}(\omega)=2 \lambda_{\rm BO}  \frac{\omega \omega_{\rm BO}^2\gamma_{\rm BO}}{(\omega_{\rm BO}^2-\omega^2)^2+\gamma_{\rm BO}^2 \omega^2} \,.
\end{equation}

\subsection{Hierarchical Equations of Motion}

In the case of a treatment with HEOM, not only thermal bath modes, but also intramolecular vibrational modes enter as bath components. In the context of HEOM a decomposition of all bath components according to an appropriate scheme is required. Here, we choose the widely-used Matsubara decomposition, where coefficients $c_k$ and frequencies $\gamma_k$ enter in the expansion of the correlation function, which is associated with an arbitrary spectral density $J(\omega)$ via
\begin{equation} \label{eq:correlation_function}
\begin{split}
C(t)=&\frac{2}{\pi} \int^{\infty}_{0} d \omega J(\omega) \\
&\times \left( \cos(\omega t) \coth \left( \frac{\omega}{2 k_B T} \right) -i \sin(\omega t) \right).
\end{split}
\end{equation}
into the series of Matsubara terms
\begin{equation} \label{eq:Matsubara_decomposition_correlation_function}
C(t)=\sum_{k} c_k \exp(-\gamma_k t).
\end{equation}
The term ``Matsubara frequency'' is commonly used for temperature-dependent frequencies $\gamma_k=2 \pi k/\beta$ with $\beta=1/k_{\rm B}T$, which appear in the Matsubara decomposition of damped bath components and stem from poles of the $\coth$-function accounting for the fluctuation-dissipation relation. In addition, so-called explicit terms, which are associated with poles of the spectral density and thus exhibit different Matsubara decomposition frequencies, enter in the series expansion. 

For the description of an undamped oscillator, i.e.\ an underdamped (Brownian) oscillator in the limit of zero damping, the correlation function can be formulated explicitly, resulting in
\begin{equation} \label{eq:coefficients_undamped_oscillators} 
\begin{split}
C_{\rm UO}(t)=&\frac{S_{\rm UO} \omega_{\rm UO}^2}{2}
\Bigg( \exp(-i \omega_{\rm UO} t) \left[ \coth\left(\frac{\beta \omega_{\rm UO}}{2}\right)+1 \right] \\
&+\exp(+i \omega_{\rm UO} t) \left[ \coth\left(\frac{\beta \omega_{\rm UO}}{2}\right)-1 \right] \Bigg).
\end{split}
\end{equation}
In this case the associated spectral density corresponds to $J_{\rm UO}(\omega)=\pi (S_{\rm UO} \omega_{\rm UO}^2/2) (\delta(\omega-\omega_{\rm UO})+\delta(\omega+\omega_{\rm UO}))$.
The respective Matsubara decomposition only contains explicit terms with the Matsubara decomposition frequencies $\gamma_{1}=i \omega_{\rm UO}$ and $\gamma_{2}=-i \omega_{\rm UO}$ 
and the attributed coefficients
\begin{eqnarray}
\label{eq:Matsubara_coefficients_UO_1}
c_{1}&=&\frac{1}{2} S_{\rm UO} \omega_{\rm UO}^2 \left( \coth\left( \frac{\beta \omega_{\rm UO}}{2} \right) +1 \right), \\
\label{eq:Matsubara_coefficients_UO_2}
c_{2}&=&\frac{1}{2} S_{\rm UO} \omega_{\rm UO}^2 \left( \coth\left( \frac{\beta \omega_{\rm UO}}{2} \right) -1 \right).
\end{eqnarray}
The derivation of HEOM has been described in detail in several publications, see e.g. Refs. \citenum{Tanimura06JPCJ,xu07_031107,schroter15_1,SeMa18_CP_129}.
Therefore, we introduce only the standard terms on the right hand side of the equations of motion for the so-called ``auxiliary density operators'' (ADOs).
The ADOs are identified by a subscript set of Matsubara indices. In the time evolution adjacent ADOs with a difference of $\pm 1$ in a single digit of their index patterns are connected to each other.
Each bath component leads to a separate segment of Matsubara indices in the index pattern \cite{LiZhBa14_JCP_134106}.
A formulation of HEOM with the rescaling introduced in Ref. \citenum{ShChNa09_JCP_164518} results in (see also Section S1 of the Supplementary Information)
\begin{equation} \label{eq:evaluation_HEOM_scheme_dimer_rescaled}
\begin{split}
\frac{\partial}{\partial t} \hat{\rho}_{\boldsymbol{n}}=&-\left( i {\cal L}_{\rm S} + \sum_{l} \sum_{k} n_{lk} \gamma_k \right) \hat{\rho}_{\boldsymbol{n}} \\
&-i \sum_{l,k}  \sqrt{(n_{lk}+1) |c_k|} \left[ \hat{B}_l^{\dagger} \hat{B}_l, \hat{\rho}_{\boldsymbol{n}^{+}_{lk}} \right] \\
&-i \sum_{l,k} \sqrt{\frac{n_{lk}}{|c_k|}} \left( c_k \hat{B}_l^{\dagger} \hat{B}_l \hat{\rho}_{\boldsymbol{n}^{-}_{lk}}
-\hat{\rho}_{\boldsymbol{n}^{-}_{lk}} \tilde{c}_k \hat{B}_l^{\dagger} \hat{B}_l \right).
\end{split}
\end{equation}
with ${\cal L}_{\rm S} \hat{\rho}_{\boldsymbol{n}}=\left[ \hat H_{\rm S}, \hat{\rho}_{\boldsymbol{n}} \right]$. The subscript $\boldsymbol{n}$ consist of a sequence of indices attributed to the terms from the Matsubara decomposition of the involved bath correlation functions.
Even though Eq.~(\ref{eq:evaluation_HEOM_scheme_dimer_rescaled}) is formulated in the localized basis, which we rely on in the present work, it can be expressed in the exciton basis as well by applying appropriate transformations to ${\cal L}_{\rm S}$, $\hat{B}_l^{\dagger} \hat{B}_l$ and to the ADOs.
In a compact formulation where the ADOs enter as the components of a Liouville space vector $\hat{\boldsymbol{\rho}}=(\hat\rho_{\bf 0},\ldots, \hat\rho_{\bf n},\ldots)$ and the terms on the right hand side are expressed by applying a matrix-valued
Liouville space operator $\lsop{L}$ to this Liouville-space vector, the hierarchical equations can be written as
\begin{equation} \label{eq:Liouville_equation_general}
\dot{\hat{\boldsymbol{\rho}}}=-i \lsop{L} \hat{\boldsymbol{\rho}}\, .
\end{equation}
The physical reduced density matrix is given by $\hat\rho_{\bf 0}(t)$.
\subsection{PT  in HEOM Space} \label{sec:pol_trans_HEOM_space}

The PT leads to a shift of the reference position with respect to selected coordinates and is also applicable in the context of HEOM, as described in the following. 
For the sake of simplicity we assume that electronic excitation of each monomer unit in a model of a molecular aggregate is coupled to a single vibrational mode. We therefore identify the index of the monomer unit with the index of the electronic state, as we already did by adopting the standard formulation of HEOM from the literature in Eq.~(\ref{eq:evaluation_HEOM_scheme_dimer_rescaled}). Further, we skip the mode index $i$ to simplify the notation.
If the shift introduced by the PT corresponds to the displacement of an excited state potential, it accounts for the situation of thermal equilibration in this state. 
While in a localized basis representation (un)damped oscillator- and/or thermal bath modes are directly coupled to the electronic excitation of specified monomer units in a description with HEOM, 
in the exciton basis the coupling of vibrational modes to the exciton states is expressed by linear combinations of contributions formulated in the localized basis. 
In the exciton basis representation with transformation coefficients $c_{\alpha l}$   between exciton state $\alpha$ and localized state $l$ the system-bath coupling Hamiltonian can be formulated as 
\begin{equation}
\label{eq:HSBexc}
 \hat{H}_{\rm SB, exc} = -\sum_{l,i} \sum_{\alpha,\beta} c_{\alpha l} c_{\beta l} \sqrt{S_{l,i}} \omega_{i} (\hat{b}_i^{\dagger}+\hat{b}_i) \hat{B}^{\dagger}_{\alpha} \hat{B}_{\beta}.
\end{equation}
Thus, electronic excitation to a single localized state leads to a coupling of all basis states from the subspace of the singly excited exciton states to the bath, where the coupling strength is multiplied by the product of transformation coefficients $c_{\alpha l} c_{\beta l}$.
Accordingly, in the hierarchical equations  $\sum_{\alpha,\beta} c_{\alpha l} c_{\beta l} \hat{B}^{\dagger}_{\alpha} \hat{B}_{\beta}$ appears instead of $\hat{B}_l^{\dagger} \hat{B}_l$ in the terms with changing index digits.
The hierarchical equations for PT with respect to vibrational coordinates coupled to $\hat{B}^{\dagger}_{\alpha} \hat{B}_{\beta}$ with a coupling strength $\sum_{l} c_{\alpha l} c_{\beta l} \sqrt{S_{l,i}}$ are formulated separately for the contibution of each localized state $l$. In other words, the following derivation performed for the local basis can readily be applied to the case of the exciton basis.

In what follows we will use the transformation operator $\lsop{D}^{\dagger}=\sum_l \lsop{D}^{\dagger}_l$, which separately accounts for the PT with respect to bath coordinates coupled to different monomer units addressed by the index $l$.
It is a Liouville space operator which acts in HEOM space, i.e.\ in the space spanned by the ADOs, on a vector $\hat{\boldsymbol{\rho}}$, yielding the polaron-transformed vector of ADOs $\lsop{D}^{\dagger} \hat{\boldsymbol{\rho}}$.
The corresponding Hilbert-space operator -- still in HEOM space -- acts as $\lsopNonCal{D}^{\dagger} \hat{\boldsymbol{\rho}} \lsopNonCal{D}$. 
However, for the following derivations it is useful not to express the PT in HEOM space, but rather in a formulation without representation in the vector space formed by the ADOs from the hierarchy. In such a formulation the shift along a selected coordinate $q_l$ up to a variable position $x_l$, which is adjusted to compensate the displacement of a selected oscillator mode coupled to electronic excitation of state $l$ \cite{ReSi96_JCP_1506}, is introduced by the shift operator  $\hat{D}(x_l)=\exp(i \hat{p}_l x_l \hat{B}_l^{\dagger} \hat{B}_l)=\exp(\hat{G}_l x_l)$ with the generator 
$\hat{G}_l=i \hat{p}_l \hat{B}_l^{\dagger} \hat{B}_l=-\sqrt{{\omega_l}/{2}}(\hat{b}_l^{\dagger}-\hat{b}_l) \hat{B}_l^{\dagger} \hat{B}_l$. 
By applying it to an arbitrary ADO from the hierarchy structure in terms of $\hat{D}^{\dagger}(x_l) \hat{\rho}_{\boldsymbol{n}} \hat{D}(x_l)$, which corresponds to the respective ADO component from the HEOM-space vector $\lsop{D}^{\dagger}(x_l) \hat{\boldsymbol{\rho}}=\lsopNonCal{D}^{\dagger}(x_l) \hat{\boldsymbol{\rho}} \lsopNonCal{D}(x_l)$, and by taking the derivative with respect to $x_l$ at $x_l=0$ one obtains
\begin{equation} \label{eq:differential_formulation_shift_operation}
\frac{\partial}{\partial x_l} \hat{\rho}_{\boldsymbol{n}}=\hat{G}_l^{\dagger} \hat{\rho}_{\boldsymbol{n}} +\hat{\rho}_{\boldsymbol{n}} \hat{G}_l
=-\hat{G}_l \hat{\rho}_{\boldsymbol{n}} +\hat{\rho}_{\boldsymbol{n}} \hat{G}_l=-\left[ \hat{G}_l, \hat{\rho}_{\boldsymbol{n}} \right].
\end{equation}
The treatment of an operator with dependence on bath coordinates in the context of HEOM, such as the momentum in the present case, has been demonstrated for non-Condon transition dipole contributions
(with dependence on the position operator of some bath mode associated with an intramolecular vibration) in Ref. \citenum{SeMa18_CP_129}.
A similar concept can be applied to account for the influence of $\hat{G}_l$.
First, we rescale the generator with $\sqrt{2 S_l \omega_l}$, so that 
$\hat{\tilde{G}}_l=\hat{G}_l \sqrt{2 S_l \omega_l}=\sqrt{S_l} \omega_l (\hat{b}_l^{\dagger}-\hat{b}_l) \hat{B}_l^{\dagger} \hat{B}_l$ can be treated on equal footing with the system-bath coupling contribution $\hat{H}_{{\rm SB},l}=\sqrt{S_l} \omega_l (\hat{b}_l^{\dagger}+\hat{b}_l) \hat{B}_l^{\dagger} \hat{B}_l$.
Furthermore, we introduce the dimensionless shift variable 
$\xi_l={x_l}/{d_l} \in [0:1]$ with  $d_l=\sqrt{{2 S_l}/{\omega_l}}$.

After drawing the connection $\hat{G}_{\xi,l}=\hat{G}_l d_l=\omega_l^{-1} \hat{\tilde{G}}_l$ between the rescaled generators, 
the next step consists in relating the mixed correlation functions defined as
${\rm Tr}_{\rm B} \{ \omega_l^{-1} \hat{\tilde{G}}_l(t) \hat{H}_{{\rm SB},l} \}=\omega_l^{-1} C_{l,1}(t) \hat{B}_l^{\dagger} \hat{B}_l$ and ${\rm Tr}_{\rm B} \{ \hat{H}_{{\rm SB},l}  \omega_l^{-1} \hat{\tilde{G}}_l(t) \}=\omega_l^{-1} C_{l,2}^{*}(t)  \hat{B}_l^{\dagger} \hat{B}_l$ to correlation functions between two system bath coupling contributions of the respective mode, namely ${\rm Tr}_{\rm B} \{ \hat{H}_{{\rm SB},l}(t) \hat{H}_{{\rm SB},l} \}=C_{l}(t) \hat{B}_l^{\dagger} \hat{B}_l$ and ${\rm Tr}_{\rm B} \{ \hat{H}_{{\rm SB},l} \hat{H}_{{\rm SB},l}(t) \}=C^{*}_{l}(t) \hat{B}_l^{\dagger} \hat{B}_l$, i.e.\ selected terms from the sum in Eq.~(\ref{eq:HSB1}).
The difference between these correlation functions consists in different Matsubara decomposition coefficients.
For the Matsubara decomposition of a mixed correlation function between a generator and a system-bath coupling component, we use a representation of the correlation function analogous to Eq.~(\ref{eq:Matsubara_decomposition_correlation_function}), but with coefficients $f_{k}$ and $\tilde{f}_{k}$ instead of $c_{k}$ and $\tilde{c}_{k}$ (with the index $k$ referring to a selected term from the Matsubara decomposition).
Further derivation steps are described in  the Supporting Information, Section S2.
Note that in general, unless an undamped oscillator is treated, the correlation function of not only a single mode, but rather of a continuum of modes described by a spectral density enters in the Matsubara decomposition. However, as no further specification of the correlation function is required in the general formulation of the hierarchical equations for PT, we postpone discussion of this aspect and immediately specify the respective hierarchical equations as
\begin{equation} \label{eq:hierarchical_equations_pol_trans}
\begin{split}
\frac{\partial}{\partial \xi_l} \hat{\rho}_{\boldsymbol{n}}&=-\left[ \hat{G}_{\xi,l,{\rm H}}, \hat{\rho}_{\boldsymbol{n}} \right] \\
&=\sum_k \sqrt{(n_{lk}+1) |f_{k}|} \left[ \hat{B}^{\dagger}_l \hat{B}_l, \hat{\rho}_{\boldsymbol{n}^{+}_{lk}} \right] \\
&+\sum_k \sqrt{\frac{n_{lk}}{|f_{k}|}} \left( f_{k} \hat{B}^{\dagger}_l \hat{B}_l \hat{\rho}_{\boldsymbol{n}^{-}_{lk}} -\tilde{f}_k \hat{\rho}_{\boldsymbol{n}^{-}_{lk}} \hat{B}^{\dagger}_l \hat{B}_l \right),
\end{split}
\end{equation}
where the subscript ${\rm H}$ in the notation $\hat{G}_{\xi,l,{\rm H}}$ denotes that the generator of the shift operator is expressed in HEOM space.
The formulation of these hierarchical equations for the PT in analogy to those for time propagation is the main result of this work.

Solving the hierarchical equations for all ADOs according to Eq.~(\ref{eq:hierarchical_equations_pol_trans}) with propagation interval from $0$ to $1$ yields $\lsop{D}^{\dagger}_l \hat{\boldsymbol{\rho}}$. In this way the equilibrium position of the respective vibrational mode is shifted by the excited state displacement. So-called variational approaches relying on the concept of PT \cite{LeMoCa12_JCP_204120,SuFuIs16_JCP_204106,YaDeJa12_JCP_024101,WaZh19_JCompChem_1097} involve a shift which does not necessarily correspond to the displacement. Such variable shift can be introduced in the framework of our approach by adjusting the upper integration boundary
in terms of multiplication with the ratio of the intended shift and the displacement.
We will continue referring to a single mode attributed to a selected site in the following discussion of how to obtain an appropriate  Matsubara decomposition with resulting coefficients $f_{k}$ and $\tilde{f}_{k}$, thereby keeping in mind that the complete PT with respect to all vibrational coordinates is obtained by applying $\lsop{D}^{\dagger}=\sum_l \lsop{D}^{\dagger}_l$ to $\hat{\boldsymbol{\rho}}$.
Note that a reformulation of Eq.~(\ref{eq:hierarchical_equations_pol_trans}) in the exciton basis is possible as well and only requires an appropriate transformation.

The correlation function between the generator of the PT in HEOM space for a selected bath mode and the contribution of the respective bath mode to the system-bath coupling can be evaluated by using properties of bosonic creation and annihilation operators and the quantum statistics of bosons.
By including also the factor ${1}/{\omega_l}$, one obtains
\begin{equation} \label{eq:correlation_function_C1}
\begin{split}
&\frac{1}{\omega_l} C_{l,1}(t)=\frac{1}{\omega_l} S_l \omega_l^2 \langle (\hat{b}_l^{\dagger}(t)-\hat{b}_l(t))(\hat{b}_l^{\dagger}+\hat{b}_l) \rangle \\
=&S_l \omega_l \left( \langle \hat{b}_l^{\dagger}(t) \hat{b}_l \rangle -\langle \hat{b}_l(t) \hat{b}_l^{\dagger} \rangle \right) \\
=&S_l \omega_l \left( \exp(i \omega_l t) \langle n_l \rangle -\exp(-i \omega_l t) \langle n_l+1 \rangle \right) \\
=&S_l \omega_l \Bigg( (\cos(\omega_l t)+i \sin(\omega_l t)) \frac{\exp(-\frac{\beta \omega_l}{2})}{\exp(\frac{\beta \omega_l}{2})-\exp(-\frac{\beta \omega_l}{2})} \\
&-(\cos(\omega_l t)-i \sin(\omega_l t)) \frac{\exp(\frac{\beta \omega_l}{2})}{\exp(\frac{\beta \omega_l}{2})-\exp(-\frac{\beta \omega_l}{2})} \Bigg) \\
=&-S_l \omega_l \left( \cos(\omega_l t)-i \coth \left( \frac{\beta \omega_l}{2} \right) \sin(\omega_l t) \right) 
\end{split}
\end{equation}
and likewise 
\begin{equation} \label{eq:correlation_function_C2}
\begin{split}
\frac{1}{\omega_l} C_{l,2}^{*}(t)&=\frac{1}{\omega_l} S_l \omega_l^2 \langle (\hat{b}_l^{\dagger}+\hat{b}_l)(\hat{b}_l^{\dagger}(t)-\hat{b}_l(t)) \rangle \\
&=S_l \omega_l \left( \cos(\omega_l t)+i \coth \left( \frac{\beta \omega_l}{2} \right) \sin(\omega_l t) \right). 
\end{split}
\end{equation}
For an undamped oscillator, which corresponds to a single mode without coupling to a thermal bath and is thus easier to treat than an underdamped (Brownian) osillator or a continuum of thermal bath modes, the Matsubara decomposition coefficients $f_{k}$ can be immediately determined from the latter equations:
By expressing the appearing $\sin$ and $\cos$ functions in terms of complex exponentials and by combining terms with the same sign in the argument of the complex exponentials
(i.e.\ with the same Matsubara decomposition frequency), one can draw a relation to the Matsubara decomposition coefficients given in Eqs.~(\ref{eq:Matsubara_coefficients_UO_1}) and (\ref{eq:Matsubara_coefficients_UO_2}), which are attributed to a correlation function between two system-bath coupling contributions. 
From Eq.~(\ref{eq:correlation_function_C1}) one obtains $f_1=-{c_1}/{\omega_{\rm UO}}$
and $f_2={c_2}/{\omega_{\rm UO}}$. Taking the complex conjugate of Eq.~(\ref{eq:correlation_function_C2}) leads to $\tilde{f}_1=-{\tilde{c}_1}/{\omega_{\rm UO}}$
and $\tilde{f}_2={\tilde{c}_2}/{\omega_{\rm UO}}$. 
In the more general case that a continuum of bath modes is described by a continous spectral density distribution, such as the Debye-Drude or Brownian spectral density from Eqs.~(\ref{eq:Debye-Drude_spectral_density}) and (\ref{eq:Brownian_oscillator_spectral_density}), one can reformulate the respective correlation functions in such way that the inverse frequency enters in their Matsubara decomposition (Supporting Information, Section S3).
Instead of Eq.~(\ref{eq:correlation_function_C1}) one then obtains
\begin{equation} \label{eq:correlation_function_C1_from_spectral_density}
\begin{split}
\bar{C}_{1,l}(t)=&-\frac{2}{\pi} \int_0^{\infty} d \omega_l \frac{J(\omega_l)}{\omega_l} \\
& \left( \cos(\omega_l t)-i \coth \left( \frac{\beta \omega_l}{2} \right) \sin(\omega_l t) \right) \\
&=-\frac{1}{\pi} \int_{-\infty}^{\infty} d \omega_l \frac{J(\omega_l)}{\omega_l} \exp(i \omega_l t) \\
&+\frac{1}{\pi} \int_{-\infty}^{\infty} d \omega_l \frac{J(\omega_l)}{\omega_l} \coth \left( \frac{\beta \omega_l}{2} \right) \exp(i \omega_l t)
\end{split}
\end{equation}
The poles of $J(\omega_l)$ and $\coth \left( {\beta \omega_l}/{2} \right)$ within the respective integration contour, which we assume to be of the general form 
$\omega_{{\rm pole},l}=i \omega_{0,l}$, are used for evaluation of the given expression, where the factor ${1}/{\omega_l}$ can be separated when the residue is determined.
Accordingly, the factor $\frac{1}{i \omega_{0,l}}$ appears, while we keep the remaining integral expressions unchanged at first, instead of evaluating their contribution to the residue.
In this way the result
\begin{equation} \label{eq:correlation_function_C1_from_spectral_density_reformulated}
\begin{split}
\bar{C}_{1,l}(t)=&-\frac{1}{i \omega_{0,l}} \left( \frac{1}{\pi} \int_{-\infty}^{\infty} d \omega_l J(\omega_l) \exp(i \omega_l t) \right. \\
& \left. -\frac{1}{\pi} \int_{-\infty}^{\infty} d \omega_l J(\omega_l) \coth \left( \frac{\beta \omega_l}{2} \right) \exp(i \omega_l t) \right),
\end{split}
\end{equation}
is obtained, which can be equivalently expressed as
\begin{equation} \label{eq:correlation_function_C1_from_spectral_density_reformulated2}
\begin{split}
\bar{C}_{1,l}(t)=&-\frac{1}{i \omega_{0,l}} \left( \frac{2}{\pi} \int_0^{\infty} d \omega_l J(\omega_l) \right. \\
& \left. \left( -i \sin(\omega_l t) +\coth \left( \frac{\beta \omega_l}{2} \right) \cos(\omega_l t) \right) \right) \\
&=-\frac{1}{i \omega_{0,l}} \left( \frac{2}{\pi} \int_0^{\infty} d \omega_l J(\omega_l) \right. \\
& \left. \left( \coth \left( \frac{\beta \omega_l}{2} \right) \cos(\omega_l t) -i \sin(\omega_l t) \right) \right).
\end{split}
\end{equation}
The integral factor from the latter expression is equivalent to the formula for calculation of a correlation function between two system-bath coupling contributions from a given spectral density.
Thus, one can use the Matsubara coefficients $c_k$ of the well-known Matsubara decomposition of such correlation functions in combination with the frequencies assigned to the $k$-th pole from the Matsubara decomposition,
$\gamma_k=\omega_{0,l}$, to determine the Matsubara coefficients $f_k$ of the correlation function $\bar{C}_{1,l}(t)$ between the generator of a shift operator and the system-bath coupling contribution of the respective bath oscillator mode. As a result one obtains $f_k=-{i c_k}/{\gamma_k}$.
Likewise, the Matsubara coefficients $\tilde{f}_k$ can be determined from $\bar{C}^{*}_{l,2}(t)$. In analogy to the derivation of Eq.~(\ref{eq:correlation_function_C1_from_spectral_density_reformulated2}) one finds
\begin{equation} \label{eq:correlation_function_C1_from_spectral_density_reformulated3}
\begin{split}
\bar{C}^{*}_{l,2}(t)=&\frac{1}{i \omega_{l,0}} \left( \frac{1}{\pi} \int_0^{\infty} d \omega_l J(\omega_l) \right. \\
& \left. \left( \coth \left( \frac{\beta \omega_l}{2} \right) \cos(\omega_l t)+i \sin(\omega_l t) \right) \right),
\end{split}
\end{equation}
which yields $\tilde{f}_k=-{i \tilde{c}_k}/{\gamma_k}$. It is immediately recognizable that in the case of an undamped oscillator the already specified Matsubara coefficients $f_1=-{c_1}/{\omega_{\rm UO}}$
and $f_2={c_2}/{\omega_{\rm UO}}$ and their counterparts $\tilde{f}_1=-{\tilde{c}_1}/{\omega_{\rm UO}}$
and $\tilde{f}_2={\tilde{c}_2}/{\omega_{\rm UO}}$ are obtained.
By using the Matsubara decomposition frequencies $\gamma_{1}=i \omega_{\rm UO}$ and $\gamma_{2}=-i \omega_{\rm UO}$ and by identifying the Matsubara-decomposed UO mode with the mode coupled to electronic excitation of monomer $l$, one obtains $\bar{C}_{1,l}(t)=\sum_{k} f_k \exp(-\gamma_k t)$ and $\bar{C}^{*}_{l,2}(t)=\sum_{k} \tilde{f}_k \exp(-\gamma_k t)$.

\subsection{Expectation Values of Vibrational Coordinates}

In Ref. \citenum{SeMa18_CP_129} the analogies between description in a vibronic basis and with HEOM were pointed out, thereby attributing the appearance of the ADOs in the HEOM approach
to the influence of bosonic creation and annihilation operators, which enter in the representation of terms with linear dependence on the bath coordinate from the system-bath coupling Hamiltonian.
Accordingly, also in the calculation of expectation values of vibrational coordinates with HEOM bosonic creation and annihilation operator lead to involvement of ADOs. 
More precisely, the expectation value of a vibrational coordinate included in a description with HEOM can be calculated from ADOs adjacent to the system density matrix $\hat{\rho}_{\boldsymbol{0}}$, as described in Ref.~\citenum{ZhLiBa12_JCP_194106}. The expectation value of a selected vibrational coordinate $q_{l}$,
which is coupled to electronic excitation of state $l$ can be obtained as
\begin{equation} \label{eq:expectation_value_vibrational_coordinate_HEOM}
\langle q_{l} \rangle_{ij}=-\frac{1}{[\hat{\rho}_{\boldsymbol{0}}]_{ij}} \left[ \sum_k \sqrt{\frac{1}{2 \omega_l}} \hat{\rho}_{\boldsymbol{0}^{+}_{lk}} \right]_{ij}.
\end{equation}
Note that a divergence can appear in the case of zero-valued matrix elements of $\hat{{\rho}}_{\boldsymbol{0}}$. In practice, such divergence can be avoided by adding a very small value to the density matrix elements to keep them from becoming equal to zero.

\subsection{Absorption and Emission Spectra}

For the calculation of absorption and emission spectra, we start from a HEOM-space vector of ADOs $\hat{\boldsymbol{\rho}}_{0,g}$ or $\hat{\boldsymbol{\rho}}_{0,e}$, where only 
the matrix elements of the system density matrix associated with populations of ground-state $g$ or excited state $e$ are non-zero initially.
Furthermore, in the case of emission a polaron transformation is applied to account for thermal equilibration in the excited state. For the calculation of the dipole-dipole correlation function
we introduce the Liouville-space dipole operators $\vec{\lsopNonCal{\mu}}_{\pm} \bullet=\left[ \hat{\vec{\mu}}_{\pm}, \bullet \right]$ with $\hat{\vec{\mu}}_{+}=\sum_m \vec{\mu}_m \hat{B}_m^{\dagger}$
and $\hat{\vec{\mu}}_{-}=\sum_m \vec{\mu}_m \hat{B}_m$. A time-dependence of the Liouville-space dipole operators is introduced via 
$\vec{\lsopNonCal{\mu}}_{\pm}(t)=\exp(i \lsop{L} t) \vec{\lsopNonCal{\mu}}_{\pm} \exp(-i \lsop{L} t)$. The dipole-dipole correlation functions of absorption and emission (with electric-field polarization $\vec{E}$) can then be written as
\begin{equation}
C_{\rm abs}(t)=\langle \langle \vec{E} \vec{\lsopNonCal{\mu}}_{-} \vec{E} \vec{\lsopNonCal{\mu}}_{+}(t) \hat{\boldsymbol{\rho}}_{0,g} \rangle \rangle
\end{equation}
and
\begin{equation}
C_{\rm em}(t)=\langle \langle (\lsop{D}^{\dagger} \hat{\boldsymbol{\rho}}_{0,e}) \vec{E} \vec{\lsopNonCal{\mu}}_{+}(t) \vec{E} \vec{\lsopNonCal{\mu}}_{-} \rangle \rangle,
\end{equation}
where $\langle \langle \bullet \rangle \rangle={\rm Tr}_{\rm S} \left\{ {\rm Tr}_{\rm B} \left\{ \bullet \right\} \right\}$ denotes the trace over both system and bath. Note that in our notation $\hat{\boldsymbol{\rho}}_{0,e}$ corresponds to an excited state population of the system density matrix, while all other ADOs are zero. Thus, this HEOM space vector describes the situation immediately after instantaneous transitions from the electronic ground state, where the bath is still equilibrated with respect to the electronic ground state. After application of PT the resulting HEOM space vector $\lsop{D}^{\dagger} \hat{\boldsymbol{\rho}}_{0,e}$ describes the vibrationally relaxed excited state.
The absorption and emission spectrum can then be calculated from the respective correlation functions via Fourier transformation:
\begin{equation}
\sigma_{\rm abs/em}(\omega)=\Re \left( \int_{0}^{\infty} \exp(i \omega t) C_{\rm abs/em}(t) \right).
\end{equation}
%

\section{Results} \label{sec:results}
%
\subsection{Validity of the PT in HEOM Space} \label{sec:demonstration_pol_trans}
%
To illustrate the effect of polaron transformation in HEOM space, we first consider separately the vibrational dynamics of an underdamped oscillator characterized by a Brownian spectral density with $S_{\rm BO}=\unit[0.5]{}$, $\omega_{\rm BO}=\unit[200]{cm^{-1}}$ and damping constant $\gamma_{\rm BO}=\unit[50]{cm^{-1}}$ 
(see Eq.~(\ref{eq:Brownian_oscillator_spectral_density})) and of a thermal bath characterized by a Debye-Drude spectral density with parameters 
$\lambda_{\rm DD}=\unit[50]{cm^{-1}}$ and $\omega_{\rm c}=\unit[50]{cm^{-1}}$ without and with the PT.
Besides the explicit terms from the Matsubara decomposition of Brownian and Debye-Drude spectral density, we took a single additional term with the lowest Matsubara frequency $\gamma_1=2 \pi/\beta$ into account, which for a temperature of $\unit[300]{K}$ has a value of $\gamma_1=\unit[1310]{cm^{-1}}$ and is thus considerably larger than the parameters $\omega_{\rm BO}=\unit[200]{cm^{-1}}$ and $\omega_{\rm c}=\unit[50]{cm^{-1}}$ entering in the respective spectral densities.
The truncation order, i.e.\ an upper bound for the values of the Matsubara index digits from the subscript index pattern of the ADOs, which may not be exceeded without the respective ADO being disregarded in the propagation, was set to a value of $\unit[14]{}$.
For calculation of the dynamics a stepsize of $\Delta t=\unit[0.0625]{fs}$ was used, for integration of the hierarchical equations related to the PT the stepsize was taken as $\Omega \Delta t$ with the same $\Delta t$ as for time propagation and with $\Omega$ corresponding to $\omega_{\rm BO}$ or $\omega_{\rm c}$, depending on whether the shift is applied to a Brownian oscillator or a thermal bath mode.

\begin{figure}[h] 
\includegraphics*[width=8cm]{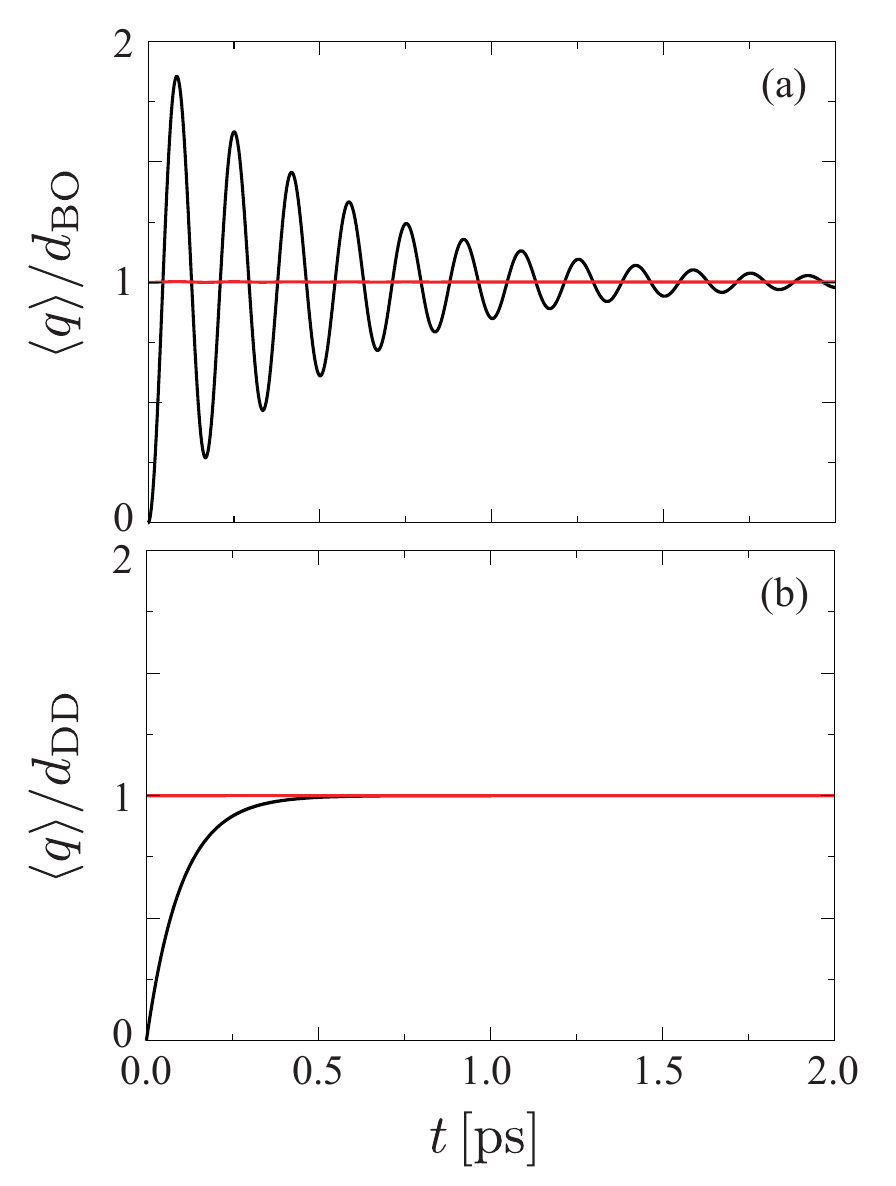}
\caption{(a) Expectation value of the vibrational coordinate of an underdamped oscillator with $S_{\rm BO}=\unit[0.5]{}$, $\omega_{\rm BO}=\unit[200]{cm^{-1}}$  and a damping constant $\gamma_{\rm BO}=\unit[50]{cm^{-1}}$ after electronic excitation, divided by the displacement in the excited state (which corresponds to $d_{\rm BO}=\sqrt{\frac{2 S_{\rm BO}}{\omega_{\rm BO}}}$).
(b) Expectation value of a bath mode attributed to a Debye-Drude spectral density ($\lambda_{\rm DD}=\unit[50]{cm^{-1}}$ and $\omega_{\rm c}=\unit[50]{cm^{-1}}$) coupled to an electronic excitation, 
divided by the displacement in the excited state (which corresponds to $d_{\rm DD}=\sqrt{2 \lambda_{\rm DD}}/\omega_{\rm c}$).
The black and red curve correspond to the results without and with PT in HEOM space, respectively.}
\label{figure1}
\end{figure}

\begin{figure}[h] 
\includegraphics*[width=8cm]{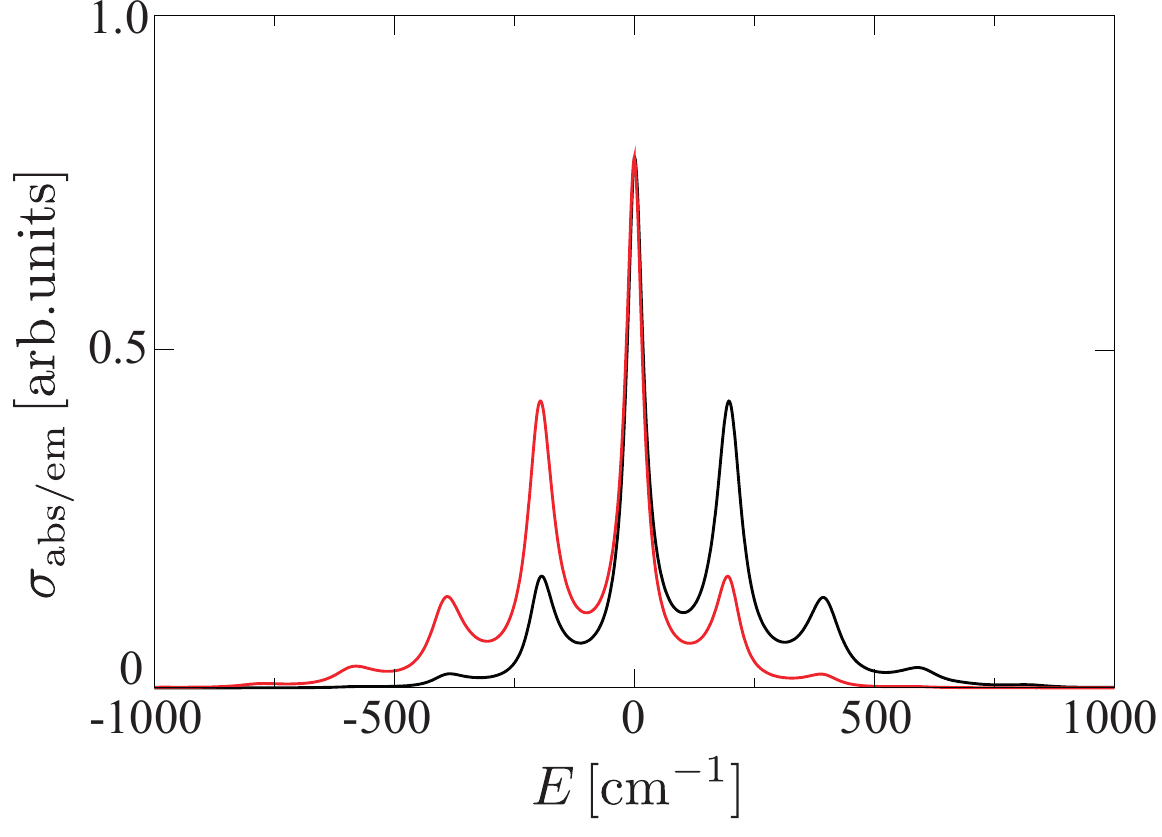}
\caption{Absorption (black line) and emission spectrum (red line) of a monomer with vibrational dynamics upon electronic excitation described by a Brownian spectral density with $S_{\rm BO}=\unit[0.5]{}$, $\omega_{\rm BO}=\unit[200]{cm^{-1}}$ and $\gamma_{\rm BO}=\unit[20]{cm^{-1}}$.}
\label{fig:absorption_emission_monomer}
\end{figure}

In Fig.~\ref{figure1} the expectation values of the respective bath coordinates,
which were calculated by adopting the approach proposed in Ref.~\citenum{ZhLiBa12_JCP_194106}, is displayed as a black line for the case of propagation without previous PT
and as a red line for the case that the PT operator $\lsop{D}^{\dagger}$ has been inserted between $\lsop{L}$ and $\hat{\boldsymbol{\rho}}$ in Eq.~(\ref{eq:Liouville_equation_general}).
In the latter case the expectation value of the selected bath coordinate remains constant at its initial value on the displayed time scale, whereas without PT damped oscillatory (Fig. \ref{figure1}a) or overdamped  (Fig. \ref{figure1}b) dynamics
with convergence towards the level of the red line appear. The results demonstrate that the present method provides the expected correct results.
%
%
Next we use the PT to account for the assumption of thermal equilibration in the excited state in the calculation of emission spectra. Monomer absorption and emission spectra are supposed to be mirror symmetric. 
For the description of the vibrational mode coupled to the electronic excitation of a monomer we use a BO spectral density with $S_{\rm BO}=\unit[0.5]{}$, 
$\omega_{\rm BO}=\unit[200]{cm^{-1}}$ $\gamma_{\rm BO}=\unit[20]{cm^{-1}}$. The resulting absorption and emission spectrum is shown in Fig.~\ref{fig:absorption_emission_monomer} as a black and red line, respectively. The expected mirror symmetry is confirmed what further supports the validity of the present approach.
%
%
%
%
In the next step will study how the PT influences the population transfer dynamics of an excitonic dimer and the expectation values of vibrational coordinates assigned to the monomer units.
\subsection{Preparation of a thermally equilibrated initial state} \label{sec:inital_state_pt}

\begin{figure}[h] 
\includegraphics*[width=8cm]{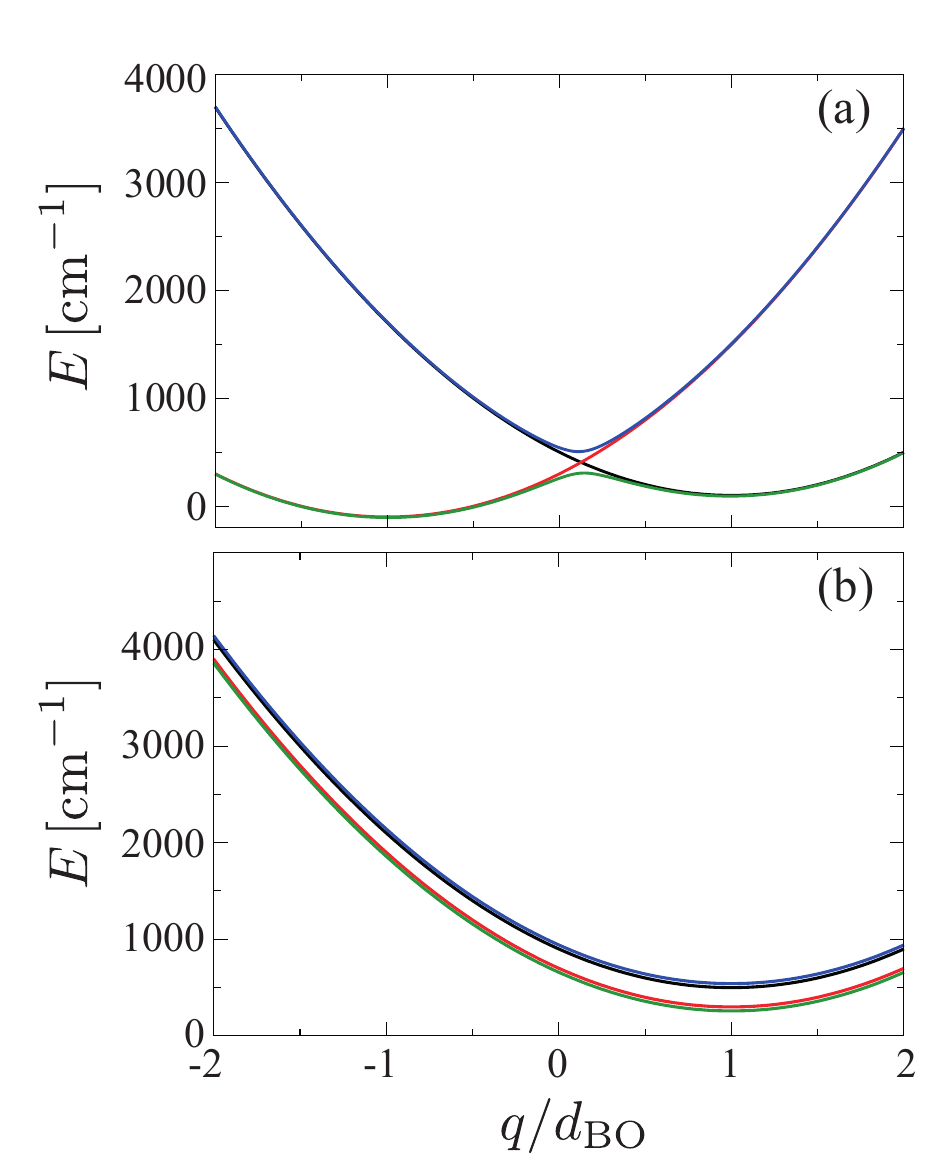}
\caption{Antidiagonal (a) and diagonal (b) cuts through excited state potentials of an excitonic dimer coupled to BO oscillators with  $S_{\rm BO}=\unit[4.0]{}$, $\omega_{\rm BO}=\unit[200]{cm^{-1}}$, and $J_{12}=\unit[100]{cm^{-1}}$. The energy offset between the donor and acceptor electronic states is $\Delta E=\unit[200]{cm^{-1}}$. The cuts through the diabatic potentials  correspond to the black and red lines, those of the cuts through the adiabatic potentials correspond to the green and the blue lines, respectively.}
\label{fig:potential_cuts}
\end{figure}

\begin{figure}[h] 
\includegraphics*[width=8cm]{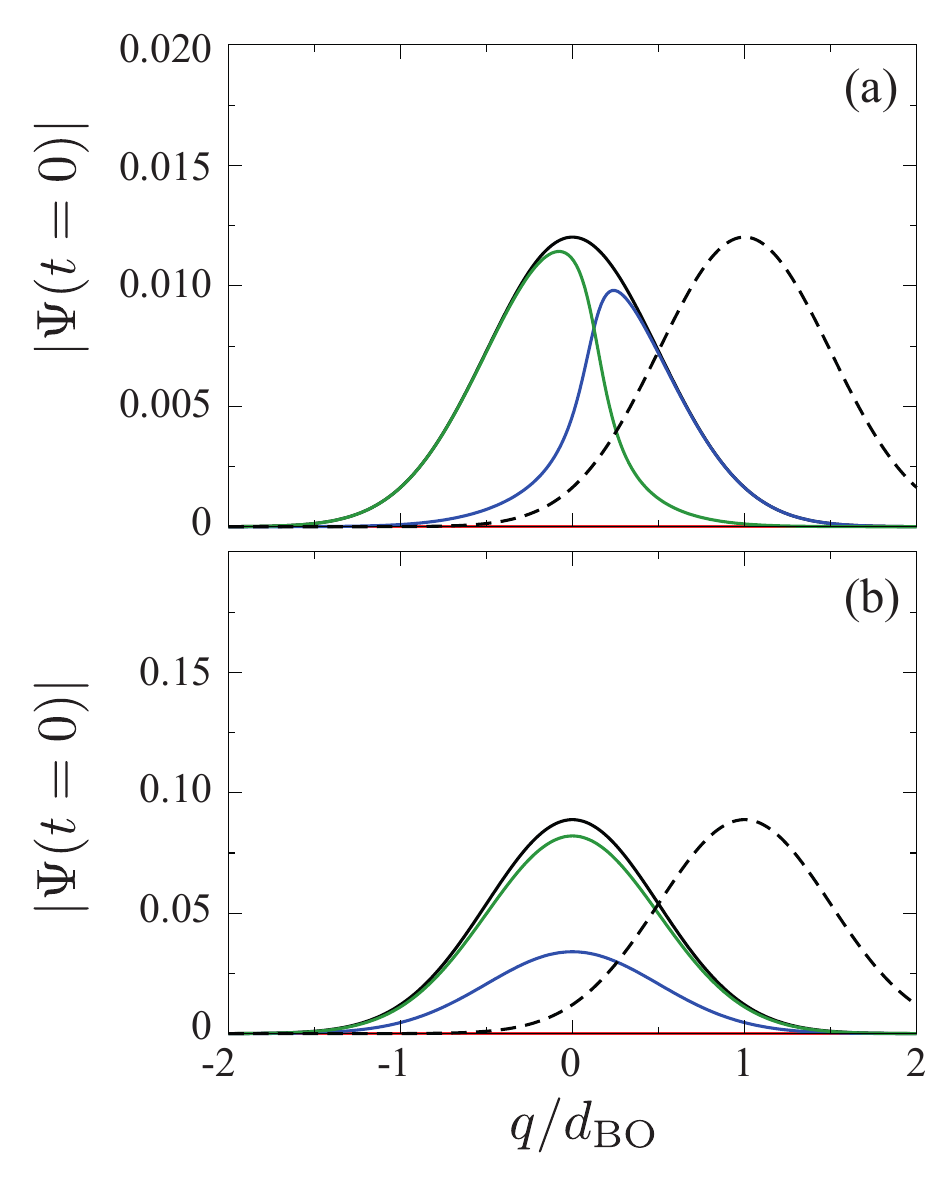}
\caption{Antidiagonal (a) and diagonal (b)  cuts through initial wavefunctions in the excited state of a dimer with parameters specified in Fig. \ref{fig:potential_cuts}. As in the corresponding illustrations of the cuts through the potentials, the black and red lines are attributed to diabatic representation, the green and the blue lines to adiabatic representation.
Furthermore, the wavefunction associated with thermal equilibration in the diabatic donor  potential is displayed as a dashed black line.}
\label{fig:wavefunctions_cuts}
\end{figure}

In the following we consider an excitonic dimer donor-acceptor system coupled to a BO with parameters chosen in such way that the monomer units with otherwise identical parameters have different electronic excitation energies with an energy gap of $\Delta E=\unit[200]{cm^{-1}}$ between the donor and acceptor. To obtain a double minimum structure in the antidiagonal cuts through the lower potential in the exciton basis with a barrier comparable to the thermal energy at $\unit[300]{K}$, we choose a large Huang-Rhys factor $S_{\rm BO}=\unit[4.0]{}$ and a relatively small excitonic coupling $J_{12}=\unit[100]{cm^{-1}}$. 
The antidiagonal and diagonal cuts through the potentials in localized and exciton basis are displayed in Fig.~\ref{fig:potential_cuts}, they will be called diabatic and adiabatic potentials, respectively.
It is assumed that the monomer unit with higher electronic excitation energy, i.e. the donor, is excited initially.
In Fig. \ref{fig:wavefunctions_cuts} we illustrate which range of the vibrational coordinates is covered by the initial excitation. To this end, we determined the wavefunctions attributed to the respective potentials in Fig. \ref{fig:potential_cuts}. Referring to the lowest vibrational wavefunction in the electronic ground state, in the case of description in the localized basis the excited state wavefunctions are obtained by multiplication with a constant factor in the framework of the Condon approximation, whereas the transformation to the exciton basis depends on the vibrational coordinates due to their influence on the eigenenergies of the localized states at the respective positions. The shapes of the wavefunctions along the antisymmetric linear combination of the vibrational coordinates and the positions of their maxima seem to be connected to the gradients of the assigned potentials of the exciton states at the absorption point which are directed toward the potential minima. 
In addition we show a shifted wavefunction which is attributed to thermal equilibrium of the vibrational mode of the monomer unit with higher electronic excitation energy as long as the excitonic coupling is disregarded.

For the calculation of the excited state dynamics and of the expectation values of the vibrational coordinates we use HEOM and study the influence of the assumption of initial thermal equilibration in the diabatic donor potential with the minimum at $q=d_{\rm BO}$. Due to the drastical increase of the Huang-Rhys factor compared to the calculations with results discussed in the monomer examples, the truncation order was increased to a value of $\unit[40]{}$ to obtain sufficient convergence. To reduce the numerical effort, only the explicit terms from the Matsubara decomposition were taken into account. Additional Matsubara terms with temperature-dependent frequencies turned out to yield a negligible contribution.

\begin{figure}[h] 
\includegraphics*[width=8cm]{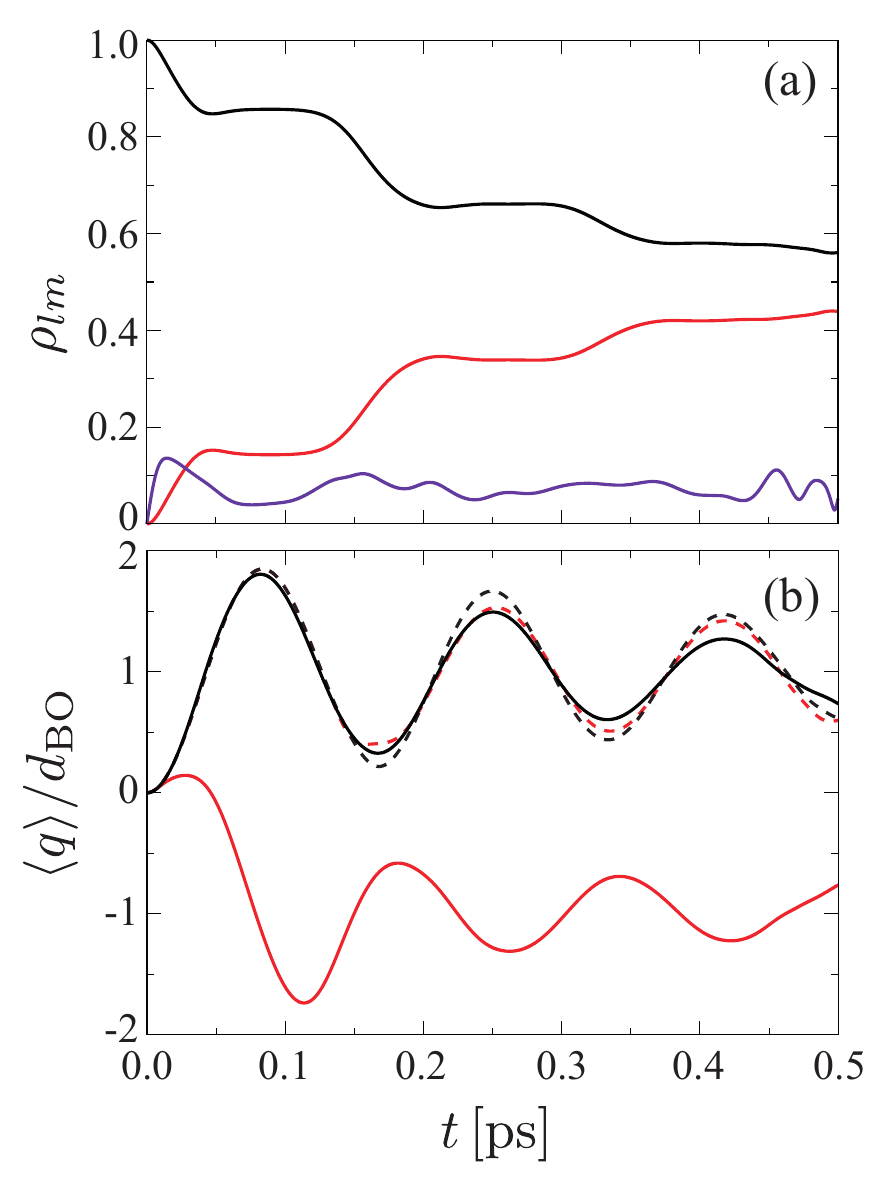}
\caption{(a) Population and coherence dynamics in the excited states the dimer of Fig. \ref{fig:potential_cuts} after vertical excitation of the donor (cf. Fig. \ref{fig:wavefunctions_cuts}).
 The black and the red line correspond to the populations of the excited state of the donor and acceptor, respectively. Furthermore, the absolut value $|\rho_{12}(t)|$ is shown as a  violet line.
 (b) Expectation values of symmetric (dashed lines) and antisymmetric (solid lines) linear combinations of vibrational coordinates. Projection to the local excited state of the donor and acceptor is indicated by black and red line color, respectively.}
\label{fig:dynamics_without_initial_PT}
\end{figure}

From a calculation in the localized basis with initial vertical excitation of only the donor we obtain the population and coherence dynamics displayed in Fig.~\ref{fig:dynamics_without_initial_PT}a) and the corresponding expectation values of vibrational coordinates displayed in Fig.~\ref{fig:dynamics_without_initial_PT}b). 
A stepwise course of the population dynamics appears, which can be attributed to the vibrational oscillations. Increased changes of the populations appear when an oscillation period of the expectation values of the vibrational coordinate is completed, i.e. upon return to the intersection between the local potentials.
Furthermore, changes in the population evolution are connected to those in the coherence evolution,
as the population transfer is mediated by the coherences. From the evolution of the expectation values of the vibrational coordinates it becomes recognizable that, starting from a value of zero immediately after excitation from the electronic ground state, oscillations of the symmetric and antisymmetric linear combinations of the vibrational coordinates around the minima of the diagonal and antidigonal cuts through the assigned potentials appear, respectively. These oscillations are damped, and as a consequence the expectation values of the vibrational coordinates tend towards the displacements of the potentials.

Next, we consider the situation where a PT has been applied after vertical excitation of the donor (cf. dashed line in  Fig. \ref{fig:wavefunctions_cuts}).
When the excited state dynamics is calculated without previous PT, the results are expected to be independent of whether representation in the localized basis is chosen or back-transformation to the localized basis is applied after calculation in the exciton basis with equivalent initial condition as in the localized basis. In practice, due to effectively decreased Huang-Rhys factors in the exciton basis \cite{SeMa18_CP_129} the convergence is better than in the localized basis for a given truncation order in the HEOM calculation, but as for the chosen parameters the calculations are sufficiently accurate, the differences are negligible.

\begin{figure}[h] 
\includegraphics*[width=8cm]{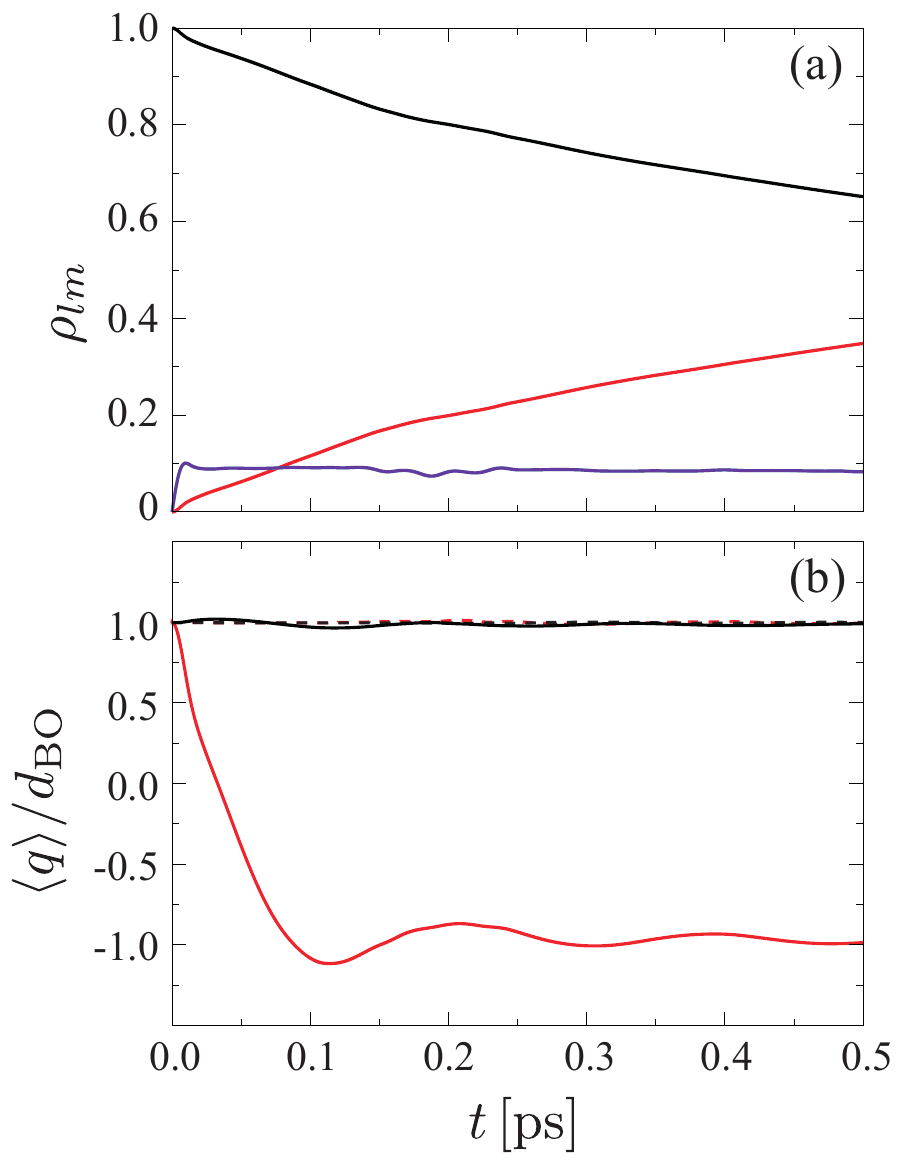}
\caption{Density matrix (a) and vibrational coordinate (b)  dynamics in analogy to Fig.~\ref{fig:dynamics_without_initial_PT}, but under the assumption of initial thermal equilibration in the donor potential, introduced via PT (color code as in Fig.~\ref{fig:dynamics_without_initial_PT}).}
\label{fig:dynamics_with_initial_PT}
\end{figure}

In Fig.~\ref{fig:dynamics_with_initial_PT}a we show the evolution of the density matrix element starting with the donor state subject to a local PT. Still population transfer appears, but as compared with Fig. \ref{fig:dynamics_without_initial_PT} there are no sudden changes and the decay has an approximately linear slope. 
The same also holds for the coherences which only exhibit changes at a very initial stage of the evolution and then remain almost constant. Different from the case without assumption of initial thermal equilibration no periodicity is recognizable anymore in the time evolution, as according to the assumption of equilibration oscillations have been completely damped and thus do not modulate the transfer process anymore. The population dynamics resembles that of a simple decay process, here  accompanied by a barrier crossing of the vibrational wavepacket.
It is not surprising that a substantial amount of population transfer takes place independent of whether PT has been applied before propagation, as in both cases non-zero off-diagonal matrix elements appear. In the case without PT they are given by the Coulomb coupling, in the case with PT they exhibit a dependence on vibrational coordinates (see Ref.~\citenum{JaChReEa08_JCP_101104}).

The expectation values of the symmetric and antisymmetric vibrational coordinates are shown in Fig.~\ref{fig:dynamics_with_initial_PT}b.
While the expectation values of the symmetric vibrational coordinate remain at their equilibrium position (corresponding to the minimum of the assigned diagonal cut through the potential of the respective electronic state), in the case of the antisymmetric coordinate the same only holds for the dynamics in the initially excited electronic state with equilibrium position determined by the minimum of the antidiagonal cut through the respective potential. Transfer to the initially unpopulated state leads to a successive change of the expectation value of the antisymmetric vibrational coordinate from the positive-signed displacement of the initial state to the negative-signed displacement of the final state associated with the minima of the antidiagonal cuts through the potentials. The oscillations which appear in the course of this transfer process are much less pronounced than those observed in the case without thermal equilibration in the initially excited localized state.
This finding indicates that the dependence of the off-diagonal couplings on bath coordinates has a minor influence and that the vibrational modes are rather sensitive to displacements of the diabatic potentials, at least for the chosen parameters.
%
%
%

In the previous examples the PT has been applied in the local electronic basis, i.e. leading to an initial wavepacket having its center at the minimum of the diabatic potential of the donor, cf. Figs. \ref{fig:potential_cuts} and \ref{fig:wavefunctions_cuts}. In the following we will perform the PT in the exciton basis, i.e. the shift $d_{\rm BO}$ towards to local minimum of the diabatic donor potential is modified to become $d_{\rm BO}c_{\alpha l} c_{\beta l}$ (here $l$ is the donor index) if a matrix element with indices $\alpha$ and $\beta$ is selected in Eq.~(\ref{eq:HSBexc}).

\begin{figure}[h] 
\includegraphics*[width=8cm]{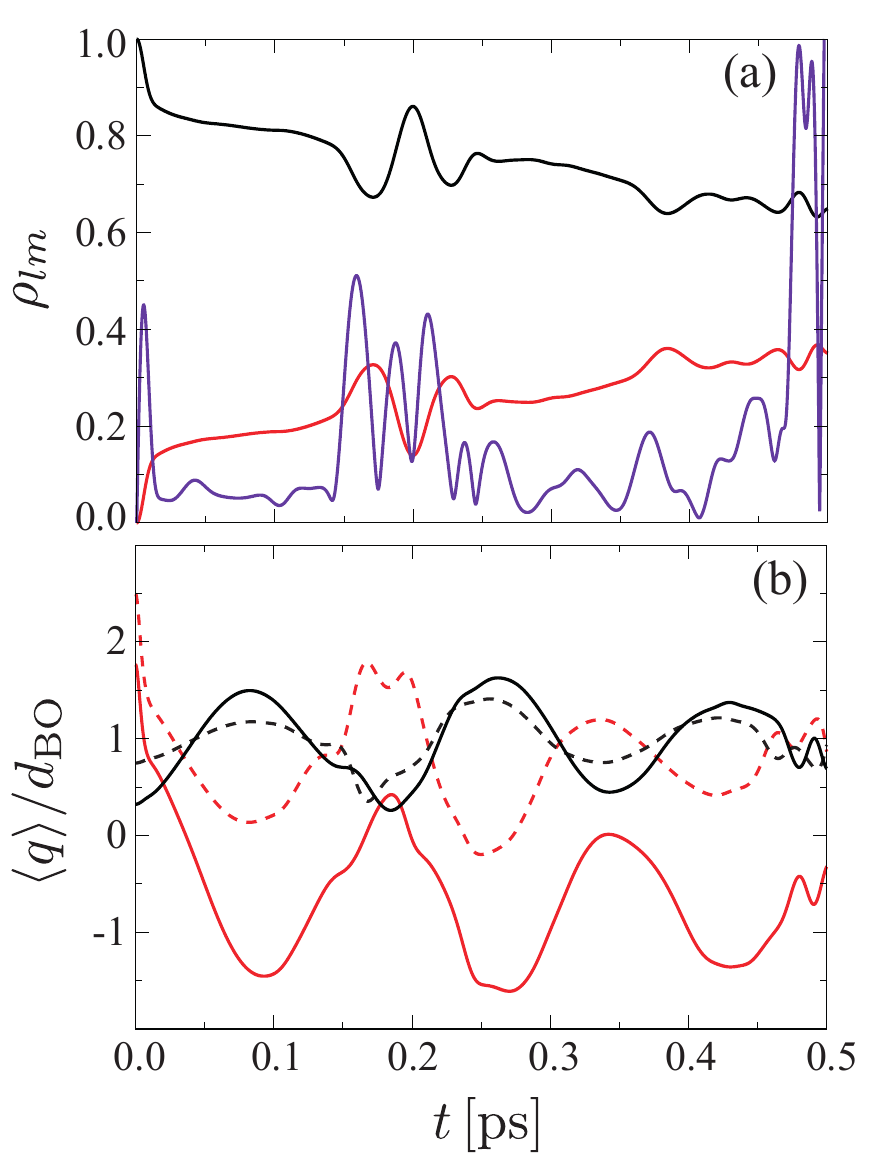}
\caption{Density matrix (a) and vibrational coordinate (b)  dynamics in analogy to Fig.~\ref{fig:dynamics_with_initial_PT}, but with an initial state which has been obtained by a PT in the exciton basis (color code as in Fig.~\ref{fig:dynamics_without_initial_PT}). Note that in the red curves in panel (b) the very initial values are somewhat biased due to the requirement of adding a small constant value to the element of the reduced density matrix in the denominator from Eq.~(\ref{eq:expectation_value_vibrational_coordinate_HEOM}) to avoid divison by zero.}
\label{fig:dynamics_with_initial_PT_exciton_basis}
\end{figure}

The resulting density matrix dynamics in the local basis is shown in Fig.~\ref{fig:dynamics_with_initial_PT_exciton_basis}a and the associated vibrational dynamics in Fig.~\ref{fig:dynamics_with_initial_PT_exciton_basis}b.
The general tendencies in the evolution of the populations are similar as in Fig.~\ref{fig:dynamics_with_initial_PT}a, i.e. in both cases the donor population decays to about 0.6 in the first 500 fs. However, in the present case the overall decay is overlaid with some oscillatory dynamics, indicating rapid population exchange during the first $\sim$30 fs and around 200 fs and 500 fs. 
 The sudden changes in the population evolution coincide with even more pronounced changes in the coherence dynamics. 
 
 The differences between Figs. \ref{fig:dynamics_with_initial_PT} and \ref{fig:dynamics_with_initial_PT_exciton_basis} can be explained by the different initial positions of the nuclei, which depend on how  'equilibration' in each basis state is accounted for by the PT. This becomes obvious from the  initial values of the black curves in Fig.~\ref{fig:dynamics_with_initial_PT_exciton_basis}b, which are not equal to $\unit[1]{}$ as in Fig. \ref{fig:dynamics_with_initial_PT}b. The actual initial value follows from the shifts introduced by the PT in the exciton basis, which exhibit a factor of $c_{\alpha l} c_{\beta l}$ as mentioned above. 
 Different from Fig.~\ref{fig:dynamics_with_initial_PT}b, where some of the expectation values of vibrational coordinate divided by their displacement remain at a value of $\unit[1]{}$, the respective curves in Fig.~\ref{fig:dynamics_with_initial_PT_exciton_basis}b oscillate around this value. In case of the projection onto the acceptor state, the oscillations around $\unit[-1]{}$ are more pronounced in Fig.~\ref{fig:dynamics_with_initial_PT_exciton_basis}b. This reflects the different equilibration in the diabatic acceptor potential.

%
%

Overall, we can conclude that neither a PT in the local basis nor in the exciton basis leads to an equilibrium exciton-vibrational state of the dimer. 

\subsection{Polaron transformation and thermal equilibration} \label{sec:pol_trans_equilibrated_state}
For a thermally equilibrated state of a molecular aggregate, such as the considered dimer, one would neither expect population transfer nor vibrational oscillations in the time evolution. Moreover, it would not play a role whether the localized basis or the exciton basis is chosen in a description with HEOM due to the non-perturbative treatment of the system-bath interaction. The PT only accounts for thermal equilibration of vibrational modes, but not for thermal equilibration with respect to electronic levels. Therefore, the question may arise whether corrections can be applied in addition to the PT, such that also the latter aspect is accounted for. Indeed, such approach has been proposed in Ref.~\citenum{XuCa16_FrontPhys_110308}, where it was suggested to determine a steady state from a QME with involvement of the PT, which was formulated in the exciton basis. In Ref.~\citenum{SeKu20_JCP} we have specified a corresponding rate kernel for calculation of transfer rates via HEOM propagations and using the PT in HEOM space. This rate kernel was formulated in the localized basis, but it can be easily adjusted for a description in the exciton basis. In the present context the formulation of the rate kernel involves representation in the polaron basis and back-transformation of the system-bath coupling components of the Hamiltonian from the interaction picture.
More details are described in Section S4 of the Supporting Information.
By applying a back-transformation from the polaron basis in every propagation step, the equilibration with respect to the electronic levels is accounted for by a description at the level of the reduced density matrix. At the same time the shifts with respect to the bath coordinates, which are introduced by the polaron transformation, are represented in HEOM space by an appropriate combination of non-zero ADOs, which corresponds to a non-equilibrium state of the bath if the distribution of the bath degrees of freedom associated with the reduced description is taken as a reference to identify the thermal equilibrium. By applying the PT such representation is recovered from the reduced description. Note that the condition for a steady state requiring a time derivative of the reduced density matrix equal to zero does not yield such steady state directly. It rather leads to an iterative procedure, from which the steady state can be determined if convergence is achieved. More details are given in the Supporting Information, Section S4. The proposed way to determine a steady state in the context of HEOM (in the framework of a second-order perturbative treatment) complements the approaches previously proposed in the literature, such as imaginary-time HEOM with integration over inverse temperature \cite{Tanimura14_JCP_044114,Ta15_JCP_144110} or self-consistent iteration \cite{ZhQiXu17_CJP_044105}.
%
\section{Conclusions} \label{sec:conclusions}
%
We have developed an approach for application of the concept of the PT in the framework of the HEOM method. To introduce a shift with respect to vibrational coordinates for compensation of an excited state displacement, we started from a transformation of the ADOs with the shift operator in differential form and expressed the influence of the generator of the respective shift in terms of connections between adjacent ADOs, resulting in hierarchical equations analogous to those for time propagation. The shift is determined by the upper integration boundary, which can also be adjusted to obtain a shift different from the excited-state equilibrium position in the framework of variational PT approaches. When a PT has been applied via integration of the respective hierarchical equtions, the introduced shift is expressed in terms of non-zero ADOs, which can be considered as components of a HEOM space vector. In a time propagation of such polaron-transformed HEOM-space vector at least the diagonal elements of the ADOs are decoupled when the introduced shift corresponds to the displacement and the system-bath interaction is thus compensated by the PT. The approach has been validated for case of dynamics in a single potential due to  an underdamped (Brownian) and an overdamped oscillator. For the underdamped oscillator we could further reproduce the mirror symmetry between absorption and emission spectra.

To study the potential and the limitations of the developed polaron transformation in HEOM space for the description of localization due to thermal equilibration in molecular aggregates, we investigated the dynamics of a dimer system with parameters chosen in such way that the lower adiabatic potential exhibits a double minimum structure with a barrier of the order of the thermal energy, which facilitates rapid transfer starting from a donor-localized state. While electronic excitation from the ground state leads to stepwise population transfer and oscillations in the expecation values of the vibrational corrdinates, an initial localization at the equilibrium position of the donor potential, modelled via PT, leads to a population evolution with almost linear slope. The expectation values of the monomer vibrarational coordinate remain at or smoothly evolve towards their expected equilibrium position in this case. 

The initial state prepared by a PT in the local basis is rather different from that obtained from a PT in the exciton basis. The reason is that although the PT can  account for thermal equilibration with respect to vibrational degrees of freedom, in the present case different approximations concerning the electronic states are involved. The situation is intuitively clear in case of the PT in the localized basis. The obtained state would be a true equilibrium state provided that the Coulomb interaction between donor and acceptor transitions is neglected. In case of the PT in the exciton basis the situation is less obvious, as the ground state vibrational density is shifted towards a position which depends on the mixing between local states in the exciton basis. 

But also the subsequent exciton-vibrational dynamics differs for the two cases and not only because of the different initial conditions. In fact, in the local basis the PT leads to a vibrational coordinate dependent Coulomb interaction, whereas in case of the PT in the exciton basis it causes a modification of the exciton-vibrational interaction Hamiltonian.

We sketched a possibility how to correct this shortcoming. With such correction it should be possible to obtain a thermally equilibrated hierarchy of ADOs, which can be taken as an initial state for calculation of, e.g., emission spectra of molecular aggregates.




\providecommand{\latin}[1]{#1}
\makeatletter
\providecommand{\doi}
  {\begingroup\let\do\@makeother\dospecials
  \catcode`\{=1 \catcode`\}=2 \doi@aux}
\providecommand{\doi@aux}[1]{\endgroup\texttt{#1}}
\makeatother
\providecommand*\mcitethebibliography{\thebibliography}
\csname @ifundefined\endcsname{endmcitethebibliography}
  {\let\endmcitethebibliography\endthebibliography}{}

\end{document}